\documentclass[twocolumn,groupaddress,10pt]{revtex4-1}

\usepackage{amsmath,amssymb,graphicx}

\begin{document}

\title{Magnetic induction tomography using an all-optical $^{87}$Rb atomic magnetometer}

\author{Arne Wickenbrock}
\author{Sarunas Jurgilas}
\author{Albert Dow}
\author{Luca Marmugi}
\author{Ferruccio Renzoni}\email{Corresponding author: f.renzoni@ucl.ac.uk}
\affiliation{Department of Physics and Astronomy, University College London, Gower Street, London WC1E 6BT, United Kingdom}

\begin{abstract}
We demonstrate magnetic induction tomography (MIT) with an all-optical atomic magnetometer. Our instrument creates a conductivity map of conductive objects. Both shape and size of the imaged samples compare very well with the actual shape and size. Given the potential of all-optical atomic magnetometers for miniaturization and extreme sensitivity, the proof-of-principle presented in this Letter opens up promising avenues in the development of instrumentation for MIT. 
\end{abstract}

\maketitle

Imaging is an essential capability in a wide range of applications, from medicine to industry and security. More than one century of development provided a variety of imaging techniques, such as  X-ray imaging, nuclear magnetic resonance (NMR) imaging, and ultrasound-based diagnostic imaging, just to name a few. Different imaging techniques rely on different properties of the object of interest, and thus provide information about different characteristics. Whenever the electrical and magnetic properties are the characteristics of interest, magnetic induction tomography (MIT) \cite{mit} is the obvious choice, as it directly provides a map of the electrical and magnetic properties of an object. Therefore, such technique is complementary to conventional magnetic imaging, and extends its range. In fact, MIT finds direct application in the detection and imaging of metallic components, e.g. for the detection of cracks or characterization of the level of corrosion. It is also a promising technique for biomedical applications, as different tissues typically present different electrical characteristics \cite{bio}. 

The ultimate performance of a MIT system depends on the magnetic field sensor used. While most of the MIT setups rely on a standard coil of wire, or an array of coils \cite{ma}, a variety of advances in different directions have been reported. 
Miniaturization can be achieved with printed circuit board (PCB) coil technology \cite{riedel}, thin film technology \cite{gatzen}, or with the use of giant magnetoresistance (GMR) sensors \cite{smith}.

In this Letter, we demonstrate MIT with all-optical atomic magnetometers. By inducing eddy currents in the object of interest, and then using an atomic magnetometer to perform position-resolved measurements (of the phase and magnitude of the magnetic field produced by these currents), we are able to produce a conductivity map of the object. Given that atomic magnetometers hold record sensitivity  \cite{budker_review} and have the potential for extreme miniaturization \cite{kitching2007a,kitching2007b,kitching2010}, this Letter paves the way for ultra-sensitive high-resolution imaging systems, using arrays of atomic magnetometers operating in a MIT modality. 

\begin{figure}[h]
	\centering
		\includegraphics[width=0.45\textwidth]{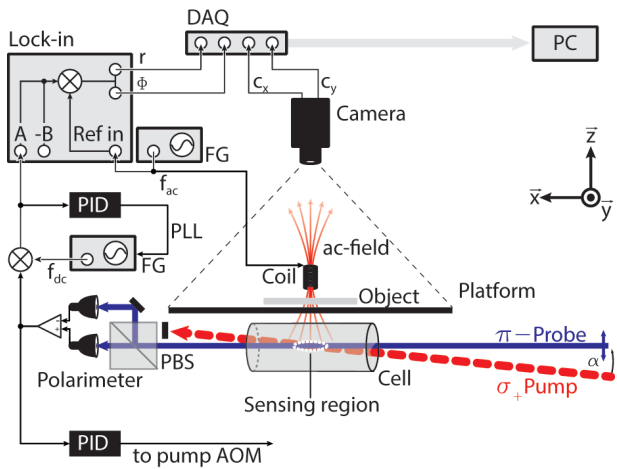}
	\caption{
Sketch of the experimental setup. A rubidium vapor cell acts as the sensor in a self-oscillating all-optical magnetometer setup. A magnetic field causes the polarization of the probe beam to oscillate at the Larmor frequency.  The oscillating polarization signal is measured with a balanced polarimeter, made of a polarizing-beam splitter cube (PBS) and two photodiodes.
An offset magnetic field applied along the $z$ axis provides a working point around $f_{dc}=$100\,kHz. An additional oscillating magnetic field is applied by modulating the current through a small coil with a function generator (FG). The coil is placed 2 mm in the $z$-direction and 75 mm in the $y$-direction with respect to the sensing region (the intersection of the pump and the probe beam). The oscillating field modulates the polarization rotation signal and induces eddy currents in a conducting object placed in its proximity. This secondary field can be detected by measuring the phase ($\Phi$) and the magnitude (r) of the signal modulation. To get a measurable component at the modulation frequency ($f_{ac}$), the polarimeter signal is multiplied with the carrier frequency ($f_{dc}$). The product signal is used as the error signal in a low bandwidth phase-locked loop (PLL), which locks a function generator to the carrier frequency, by means of a proportional-integral-derivative (PID) controller. The unfiltered signal is fed to a lock-in amplifier, with the driving frequency as the reference signal (i.e., Ref in).
 To create images, the phase and magnitude of the modulation signal are recorded with a data acquisition device (DAQ) and a personal computer (PC), while varying the center position (C$_x$ and C$_y$) of the object, which is detected by a CCD camera.
 }.
	\label{fig:Figure1}
\end{figure}

\begin{figure*}
	\centering
	\includegraphics{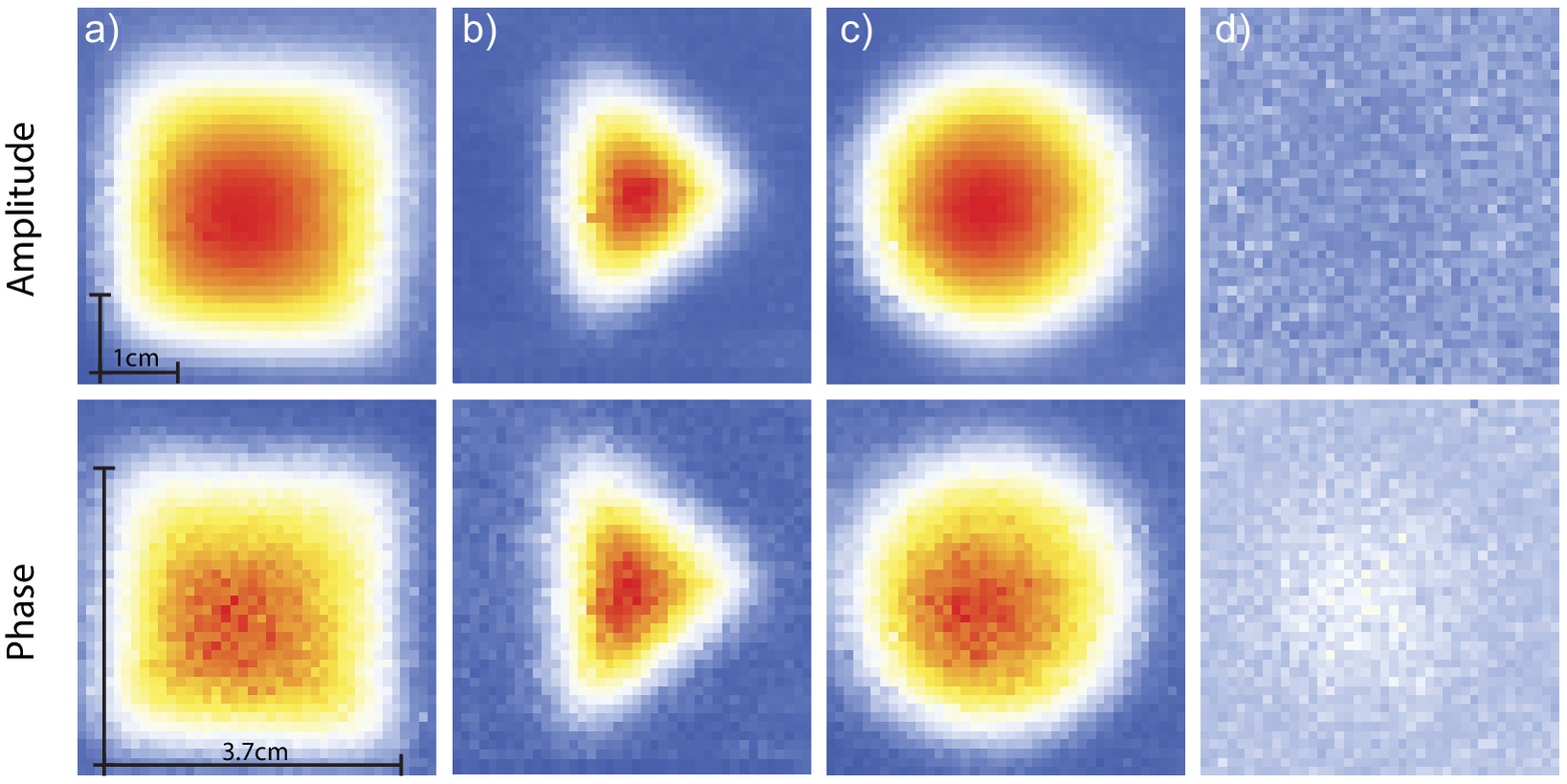}
	\caption{Normalized magnetic induction tomography: (a-c) different objects, (d) an example of the acquisition error, multiplied by a factor of 3 (phase) and 20 (amplitude), to be visible with the respective color coding. 
The first row shows the position resolved normalized amplitude of the ac magnetic field signal as detected by the lock-in amplifier. The second row shows the corresponding normalized phase data.  Both amplitude and phase variation depend on the position of the object, with respect to the driving coil:  (a) data for a $37$ x $37\text{mm}^2$ square,
(b) an isoscele triangle with one side of 37\,mm and two sides of 30\,mm, (c) a disk with 37\,mm diameter, (d) amplified acquisition error for the disk data. All objects were made from 2\,mm thick aluminium sheets.}
	\label{fig:Figure2}
\end{figure*}

The experimental apparatus is shown schematically in Fig. \ref{fig:Figure1}. The object of interest is placed on a horizontal flat nonconductive support, and can be moved manually. The atomic magnetometer for magnetic field sensing is under the support. The sensor is a 5 cm long vapor cell filled with the naturally occurring mixture of $^{85}$Rb and $^{87}$Rb. The cell is coated with polydimethylsiloxane (PDMS), and filled with 5 Torr of argon gas. It is heated to 70$^\circ$C to increase the vapor density. The magnetometer operates in self-oscillating mode \cite{schwindt,belfi}. The detailed description of our magnetometer is reported in Ref.~\cite{arne}, and we summarize in this Letter only the essential information. 
A circular polarized laser beam, tuned  80\,MHz to the red of the $F=2 \to F=3$ D$_2$ line transition of $^{87}{\rm Rb}$, is used to optically pump the atoms. A linearly polarized probe laser, detuned by 360 MHz in the blue of the $F=2 \to F=3$ D$_2$ line transition, is used to probe the atomic vapor. To operate the magnetometer in self-oscillating mode, the pump laser intensity is modulated by the polarization rotation signal of the blue detuned probe laser. For appropriate gain settings, this causes the system to oscillate at a frequency directly related to the Larmor precession frequency. The oscillation frequency of the magnetometer can thus be used to directly measure the magnetic field. In this Letter, we are interested in measuring rapidly alternating magnetic fields, superimposed to a static magnetic field. This can easily be done as the ac magnetic field results in sidebands in the frequency spectrum of the oscillating polarization signal. 
The strength of the dc field, which is directly related to the Larmor frequency $f_{\rm dc}$, and the applied ac field frequency $f_{\rm ac}$ are chosen depending on the application. In particular, they are determined by the material properties of the samples, and the required penetrating power. Once the ac frequency is selected, the static field is chosen so that  the self-oscillating frequency of the magnetometer depends linearly on the Larmor frequency in the range $[f_{\rm dc}-f_{\rm ac};f_{\rm dc} + f_{\rm ac}]$.
 In all the measurements presented in this Letter, both static (offset) and ac magnetic fields are applied along the $z$ axis. For the proof-of-principle presented in this Letter, we  consider metallic objects with large conductivity. Without specific requirements of penetrating power, imaging is possible over a wide range of frequencies, from a few Hz to several MHz. We arbitrarily set 10 kHz as the frequency of our ac magnetic field. The offset magnetic field is then set to around 100 kHz. We verified that the frequency of self-oscillation displays a linear dependence with the applied magnetic field in the range $80-120$ kHz. This guarantees a linear response for the ac magnetic fields, with a frequency of 10 kHz used in the research reported in this Letter. 

The alternating magnetic field is produced by a small coil of 4.6\,mm diameter, 10\,mm length, and an inductance of 100\,$\mu$H.  The coil is placed 2 mm above the sensing region determined by the intersection of the laser beams. It is also displaced by 75 mm in the $y$-direction (as defined in Fig. \ref{fig:Figure1}). A FG is used to apply a 10\,kHz current modulation directly to this coil. This creates an oscillating magnetic field along the $z$-direction with around 0.5 G amplitude at the position of the object. The resulting field variation amplitude as measured at the sensing region is about 2 mG.

The oscillating polarization signal of the probe beam contains a component due to the applied ac field. The self-oscillating signal becomes frequency modulated. 
In frequency space, this produces sidebands around the dc field frequency, i.e., the carrier, at $f_{\rm dc} \pm f_{\rm ac}$. To access the amplitude and phase of this oscillating signal as measured by the magnetometer, we multiply the polarimeter output (including the sidebands) with a sinusoidal voltage, oscillating at the carrier frequency only. This is achieved by a low bandwidth phase lock of a FG to $f_{\rm dc}$. The product signal contains components oscillating at the modulation frequency, which are then detected with a dual-phase lock-in amplifier referenced to the driving signal.
The lock-in amplifier then allows us to measure the amplitude and phase of the total oscillating field, i.e., the superposition of the primary driving field and the secondary field resulting from eddy currents in the sample.

In general, the amplitude of the oscillating magnetic field and its phase lag are determined by the material properties of the object: electrical conductivity $\sigma$, relative permittivity  $\epsilon_r$ and relative permeability $\mu_r$ \cite{griffiths07}. In the proof-of-principle presented in this Letter, we will consider objects with large conductivity, but small relative permeability and permittivity. We can thus assume that both amplitude of the secondary field, as well as its phase lag, are determined by the conductivity.

For the tomographic measurements, the object of interest is placed on the nonconductive support. In order to take spatially resolved measurements, the object is moved manually to different positions, with respect to the measuring apparatus, by means of a micrometric translational stage; consequently, manual positioning does not affect the imaging procedure. For each position, $10^3$ samples of the phase and amplitude signals are measured in a 20\,ms interval. The position of the object is determined by a CCD camera placed above the non-conductive support. The mean value and the standard deviation of the $10^3$ measurements of phase and amplitude is computed and, together with the sample position data, acquired by the data acquisition system.

As a proof-of-principle for MIT with all-optical atomic magnetometers, we imaged three differently shaped objects: a $37$ mm x $37$ mm square, a disk with 37\,mm diameter and an isoscele triangle (with one side of 37\,mm and two sides of 30\,mm). All the objects were made of 2\,mm thick aluminium sheets.  We notice that at the used frequency of 10 kHz, the skin depth of aluminium is 0.82 mm, less than half of the sample thickness. Our instrument therefore images the surface of the object, in which eddy currents circulate.
Figure \ref{fig:Figure2} shows the results of our experiment. Data for the variation of the ac magnetic field amplitude induced by the presence of the objects were produced
by normalizing the acquired data to the maximum amplitude and phase change, and subtracting a constant background level. In this way, the amplitude and phase lag reported in Fig. \ref{fig:Figure2} are the ones determined by the presence of the object. Images are therefore produced with a 2D spatial representation of normalized amplitude and phase data.
The instrument is clearly able to resolve shapes, as demonstrated by the well distinguished images of the three different objects in Fig.  \ref{fig:Figure2}. Also, the images produced via the amplitude of the signal and the phase lag are qualitatively very similar. This was expected, as for aluminium the effect produced by the conductivity is dominant, and the permittivity and permeability of the material can be neglected. Thus both tomographic images, obtained via position-resolved amplitude and phase measurements, are determined by the conductivity. 

\begin{figure}[htbp]
\centering
\includegraphics[width=0.45\textwidth]{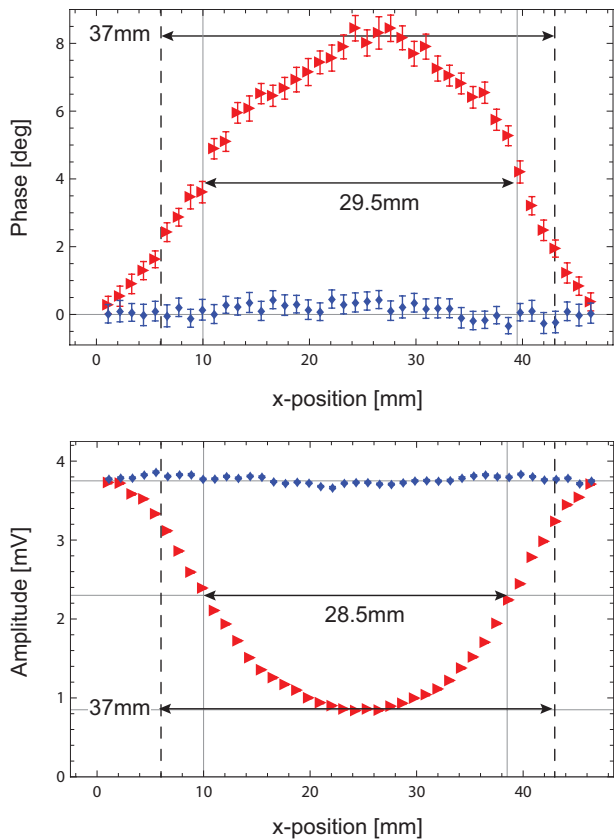}
\caption{Cross section through the center of the disk data (red triangles), compared to the background (blue filled diamonds). The vertical dashed lines indicate the  extension of the disk. The vertical grey lines mark the FWHM size of the object. The horizontal solid grey lines mark the extreme and intermediate points in the signal used to determine the FWHM.
}
\label{fig:Figure3}
\end{figure}

Additional information on the imaging capability of our instrument can be obtained by inspecting a cross section of the image data, which is reported in Fig. \ref{fig:Figure3} for the case of the aluminium disk. In Fig. \ref{fig:Figure3}, raw data for the phase and amplitude are compared to the background data acquired with no object present. 
It is important to examine how the size of the object, as measured from the magnetic image, compares with the real dimension. Figure \ref{fig:Figure3} shows that the magnetic image extends over a radius which corresponds, within the error, to the actual radius of the aluminium disk. Given that the sensor is 5\,cm long, such an agreement was not obvious {\it a priori}. However, as the driving field is well-localized, eddy currents are only induced in the portion of the object closest to the coil. Hence, the good resolving power of the instrument.  To be more quantitative, we take the FWHM (of the phase and amplitude profiles) as measurement of the size of the disk. As from Fig. \ref{fig:Figure3}, the FWHM is 29.5 mm for the phase measurements, and 28.5 mm for the amplitude measurements. The two widths coincide within the experimental error,  as expected  due to the large conductivity of the object. 
The reduced measured radius, as derived from the FWHM, is due to the distribution of eddy currents on the surface of the object, so that the field and phase at the center are larger than at the edges. A more accurate reconstruction of the object from our measurement can be made by standard inverse problem techniques \cite{inverse}.
 
In conclusion, we demonstrated MIT with an all-optical atomic magnetometer. Our instrument creates a conductivity map of conductive objects, and their shape and size are very well-distinguishable.  With respect to the standard approach based on a pick-up coil, our technique has the significant advantage of using atomic magnetometers that have been shown to be more sensitive than standard pick-up coils, for frequencies below 50 MHz \cite{savukov}.  The specific settings and the imaging layout should be designed in view of the field of application; however, given the potential for miniaturization and extreme sensitivity, the demonstrated setup offers potentially large improvements to current MIT instruments.

\section*{Acknowledgment}

This Letter was supported by EPSRC (Grant No. EP/H049231/1),  and a Marie Curie International Research Staff Exchange Scheme Fellowship 
within the 7th European Community Framework Program. L. M. thanks the Foundation Angelo della Riccia for a scholarship. 
We would like to thank Emilio Mariotti for providing the rubidium cell used in the research reported in this Letter.


\begin{thebibliography}{99}
\bibitem{mit}
H. Griffiths, Meas. Sci. Technol. {\bf 12}, 1126 (2001)
\bibitem{bio}
M. Zolgharni, H. Griffiths and P. D. Ledger, Physiol. Meas. {\bf 31}, S111 (2010).
\bibitem{ma}
L. Ma, H.-Y. Wei, and M. Soleimani,
Progress in Electromagnetic Research M {\bf 23}, 65 (2012).
\bibitem{riedel}
C. H. Riedel, M. Keppelen, S. Nani, R. D. Merges, and O. D\"ossel, Physiol. Meas. {\bf 25}, 403 (2004).
\bibitem{gatzen}
H. H. Gatzen, E. Andreeva, and H. Iswahjudi
IEEE Trans. on Magn. {\bf 38} 3368 (2002).
\bibitem{smith}
C. H. Smith, R. W. Schneider, T. Dogaru, and S. T. Smith,
AIP Conf. Proc. {\bf 700}, 406 (2004).
\bibitem{budker_review}
D. Budker, and M. V. Romalis, Nat. Phys. {\bf 3}, 227 (2007).
\bibitem{kitching2007a}
V. Shah, S. Knappe, P. D. D. Schwindt, and J. Kitching, Nat. Photonics {\bf 11}, 649 (2007).
\bibitem{kitching2007b}
P. D. D. Schwindt, B. Lindseth, S. Knappe, V. Shah, J. Kitching, and L.-A. Liew,  Appl. Phys. Lett. {\bf 90}, 081102 (2007).
\bibitem{kitching2010}
W. C. Griffith, S. Knappe, and J. Kitching, Opt. Express {\bf 18}, 27167 (2010).
\bibitem{schwindt}
P. D. D.~Schwindt, L.~Hollberg, and J.~Kitching,
Rev. Sci. Instrum. {\bf 76}, 126103 (2005).
\bibitem{belfi}
J.~Belfi, G.~Bevilacqua, V.~Biancalana, S.~Cartaleva, Y.~Dancheva, K.~Khanbekyan and L.~Moi, 
J. Opt. Soc. Am. B {\bf 26}, 910 (2009).
\bibitem{arne}
A. Wickenbrock, F. Tricot, and F. Renzoni, Appl. Phys. Lett. {\bf 103}, 243503 (2013).
\bibitem{griffiths07}
H. Griffiths, W. Gough, S. Watson, and R. Williams, Physiol. Meas. {\bf 28}, S301 (2007).
\bibitem{inverse}
R. Merwa, K. Hollaus, P. Brunner and H. Scharfetter,
Physiol. Meas. {\bf 26}, S241 (2005).
\bibitem{savukov}
J. M. Savukov, S.J. Seltzer, and M.V. Romalis, J. Magn. Res. {\bf 185}, 214 (2007).
\end{thebibliography}
\end{document}